\documentclass[twocolumn,preprintnumbers,aps]{revtex4} 
\usepackage{graphicx} 
\begin{document}

\title{ Atomistic theory of electronic and optical properties of InAs/InP self-assembled quantum dots on patterned substrates }

\author{ Weidong Sheng } \email{ weidong.sheng@nrc-cnrc.gc.ca }

\author{ Pawel Hawrylak }

\affiliation{ Institute for Microstructural Sciences, National Research Council of Canada, Ottawa, ON K1A 0R6, Canada }

\begin{abstract} 
We report on a atomistic theory of electronic structure and optical properties of a single InAs quantum dot
grown on InP patterned substrate. The spatial positioning of individual dots using InP nano-templates 
results in a quantum dot embedded in InP pyramid. The strain distribution of a quantum dot in InP pyramid 
is calculated
using the continuum elasticity theory. The electron and valence hole single-particle states are calculated
using atomistic effective-bond-orbital model with second nearest-neighbor interactions, coupled to
strain via Bir-Pikus Hamiltonian. The optical properties are determined by solving many-exciton Hamiltonian
for interacting electron and hole complexes using the configuration-interaction method. The effect of
positioning of quantum dots using nanotemplate on their optical spectra is determined by a comparison with
dots on unpatterned substrates, and with experimental results. The possibility of tuning the quantum dot
properties with varying the nano-template is explored.
% 73.21.La Electron states in nanoscale systems - Quantum dots 
% 78.67.Hc Optical properties of nanoscale materials and structures - Quantum dots 
% 71.15.-m Methods of electronic structure calculations 
\end{abstract} 

\maketitle

\section{Introduction}

There is currently significant interest in semiconductor self-assembled quantum dots (SAQDs) \cite{QD:Jacak,
QD:Bimberg, QD:Hawrylak} due to their excellent electronic and optical properties. The quality of optical
properties is primarily due to the very clean self-organized Stransky-Krastanow growth process during
molecular beam epitaxy. The downside of this process is the random spatial distribution of quantum dots. In
order to combine single quantum dots with cavities, gates, or magnetic ions for their increased
functionality one needs to position quantum dots by growth on patterned substrates \cite{QD:PS}.
 
The patterned substrate growth is both a challenge and opportunity. It is a challenge to assure that
patterning of the substrate does not destroy the high quality of single quantum dots, and it is an
opportunity to use patterning to control their electronic properties. Recently Williams et al
\cite{EXP:Williams, EXP:Chithrani} reported high quality emission spectra close to 1.55 $\mu m$ of single
InAs quantum dots grown on InP pyramidal nanotemplate. This opens up the possibility of integration of
single quantum dots into optical cavities \cite{EXP:Dalacu}, and hence reliable fabrication of single photon
sources \cite{EXP:Kim, EXP:Benson} suitable for long distance transmission in optical fiber. The theoretical
understanding of the effect of patterning on optical spectra is unknown, and we present here atomistic
theory of electronic and optical properties of single InAs quantum dot grown on pyramidal InP nanotemplates.

\begin{figure} \vspace{5mm}
  \includegraphics[width=3in]{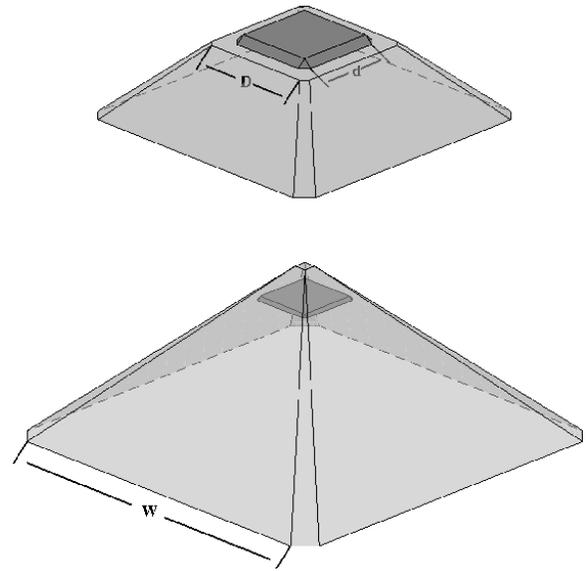}\\
  \caption{ Single InAs quantum dot on a pyramidal InP nanotemplate (a) before overgrowth, 
enlarged view of top portion of the structure and (b) after overgrowth. }
\label{dotonps}
\end{figure}

Atomistic calculation of the electronic properties of InAs dots on InP presents a challenge: due to small
lattice mismatch between InAs and InP ($3\%$), the InAs dots are significantly larger then InAs dots on
GaAs. The larger dot size would suggest the applicability of the ${\bf k}\cdot{\bf p}$ method, unfortunately
it suffers from the unphysical level crossing as pointed out by Holma et al \cite{KP:Holma}. Hence atomistic
method, which gives proper bulk band dispersion, is needed for InAs/InP dots. In this work, the electron and
valence hole single particle states are calculated using atomistic effective-bond-orbital model (EBOM)
\cite{EBOM:Chang, EBOM:Sun} with second nearest-neighbor interactions, coupled to separately calculated
strain distribution via Bir-Pikus Hamiltonian \cite{DP:Bir}. The optical properties of InAs dots embedded in
InP pyramids are calculated by solving the many-exciton Hamiltonian for interacting electron and hole
complexes using the configuration-interaction method. The effect of positioning of quantum dots using
nanotemplate on their optical spectra is determined by a comparison with dots on unpatterned substrates, and
with experimental results. The possibility of tuning the optical properties with the shape of the
nanotemplate is demonstrated.

\section{Model of InAs quantum dot on nanotemplate}

In the work of Williams et al \cite{EXP:Williams, EXP:Chithrani} the positioning of InAs dots is achieved 
by first the growth of InP nano-template with controlled size and shape, followed by the growth of a 
single InAs dot, as shown in Fig. \ref{dotonps}. The quantum dot is covered by continuing the growth of 
InP, with the final result of a single InAs dot embedded in an InP pyramid. There is experimental 
evidence \cite{EXP:Williams} that the shape and size of the template determines the geometry of the dot. 
Hence for a rectangular template, including facets associated with [100] and [111] crystallographic axis 
shown in Fig. \ref{dotonps}, we model the dot as a rectangular InAs box, with truncations around its four 
corners.

In this calculation, we chose the size of the template as $D$=46 nm, the lateral size of the dot as 
$d$=36 nm, with height of $h$=2.35 nm \cite{EXP:McCaffrey}. After overgrowth, the dot ends up embedded 18 
nm below the top of the pyramid, as shown in the lower panel of Fig. \ref{dotonps}. The size and the 
shape of the nanotemplate (denoted by $W$ in Fig. \ref{dotonps}) can be varied. The pyramid fabricated in 
Refs. \cite{EXP:Williams, EXP:Chithrani} had $W$ = 400 nm. In this exploratory example, we chose a
smaller size of the pyramidal template, $W$ = 206 nm with height $h$ = 103 nm to reduce significant 
computational effort.

\section{Atomistic tight-binding approach to electronic structure}

Electronic structure of InAs/InP dots has been calculated in the past using the quasicontinuous 
eight-band ${\bf k}\cdot{\bf p}$ method \cite{EXP:Pettersson2000}. For atomistic calculations, the 
empirical tight-binding method (ETBM) \cite{TB:Bryant} is a natural choice.  However, considering that 
InAs/InP dots are generally much larger than InAs/GaAs dots due to smaller lattice mismatch, we chose 
EBOM due to its less requirement of computational resources and a compatibility with the $k*p$ method. 
EBOM is a $sp^3$ tight-binding method in which the full symmetry of zinc-blende lattice is reduced to 
that of a $fcc$ lattice. The one electron tight-binding Hamiltonian describes an electron hopping from 
atomic orbital $\alpha$ at position ${\bf R}$ to orbital $\alpha'$ at position ${\bf R'}$:
\begin{equation}
H_{tb} =  \sum_{  {\bf R},\alpha,{\bf R^\prime},\alpha'}    
H (       {\bf R}, \alpha ;    {\bf R^\prime},\alpha'  )
 c_{  {\bf R}, \alpha } ^+ c_{   {\bf R^\prime},\alpha'    }
  ,
\label{tb-hamiltonian}
\end{equation}
where $ c_{{\bf R}, \alpha } ^+ (  c_{{\bf R^\prime},\alpha' })$ are creation ( annihilation ) operators.
The hopping matrix elements and site energies for $s$ and $p$ orbitals, extended to include second
nearest-neighbor interactions, are given in real space by
\begin{eqnarray}
H({{\bf R}s,{\bf R^\prime}s}) &=& E_{ss}^{000} \delta_{{\bf R},{\bf R^\prime}} + E_{ss}^{110} \delta_{{\bf R}-{\bf 
R^\prime},\tau} + E_{ss}^{200} \delta_{{\bf R}-{\bf R^\prime},\sigma} , \cr
H({{\bf R}p,{\bf R^\prime}p} ) &=& 
\delta_{{\bf R}-{\bf R^\prime},\tau} \bigl[ E_{xx}^{110}\tau_p^2 + E_{xx}^{011}(1-\tau_p^2) \bigl] \cr
&+& \delta_{{\bf R}-{\bf R^\prime},\sigma} \bigl[ E_{xx}^{200}\sigma_p^2 + E_{xx}^{002}(1-\sigma_p^2) \bigl] \cr 
&+& E_{xx}^{000} \delta_{{\bf R},{\bf R^\prime}} , \cr
H({{\bf R}p,{\bf R^\prime}p^\prime}) &=& E_{xy}^{110} \tau_p \tau_{p^\prime} \delta_{{\bf R}-{\bf R^\prime},\tau} , \cr
H ( {{\bf R}s,{\bf R^\prime}p} ) &=& E_{sx}^{110} \tau_p \delta_{{\bf R}-{\bf R^\prime},\tau} ,
\label{EBOM}
\end{eqnarray}
where $\tau$ and $\sigma$ give positions of the nearest and second-nearest neighbors, respectively,
\begin{eqnarray}
\tau   &=& \frac{a}{2} \bigl[ (\pm1,\pm1,0), (\pm1,0,\pm1), (0,\pm1,\pm1) \bigl] , \cr
\sigma &=& a \bigl[ (\pm1,0,0), (0,\pm1,0), (0,0,\pm1) \bigl] ,
\end{eqnarray}
$a$ is the lattice constant. 

\begin{figure}
\vspace{5mm}
  \includegraphics[height=2.5in]{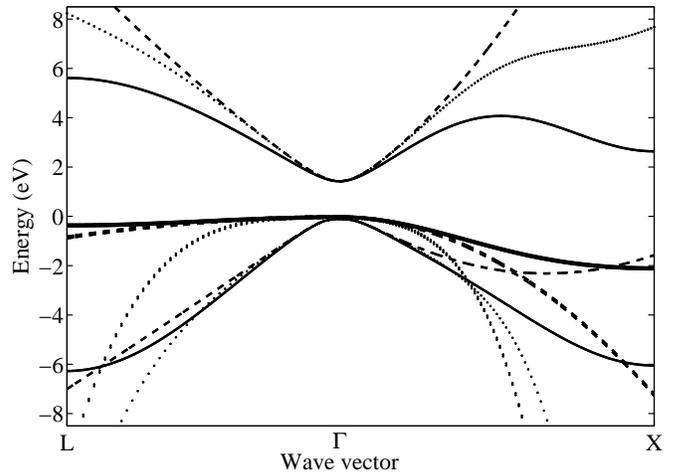}\\
  \caption{ Band structure of InP described by the eight-band ${\bf k}\cdot{\bf p}$ method (dotted lines) and its
modified version (dash lines, see text), compared with that by the tight-binding-like EBOM (solid lines). }
\label{ebands}
\end{figure}

There are ten fitting parameters, $E_{ss}^{000}$, $E_{ss}^{110}$, $E_{ss}^{200}$, $E_{sx}^{110}$, 
$E_{xx}^{000}$, $E_{xx}^{110}$, $E_{xx}^{011}$, $E_{xx}^{200}$, $E_{xx}^{002}$, $E_{xy}^{110}$ in EBOM. 
The two-center approximation \cite{EBOM:Loehr} introduces an additional constraint, 
$E_{xx}^{110}-E_{xy}^{110}=E_{xx}^{011}$. With second nearest-neighbor interactions included 
\cite{EBOM:Meyer}, effective masses of electrons and holes at the $\Gamma$ point \cite{KP:Holma, 
KP:InAs/GaAs}, conduction and valence band edges at both the $\Gamma$ and X points \cite{TB:Klimeck} can 
be exactly fitted within EBOM. With increasing $k$, it can reproduce correct band dispersion along the 
$\Gamma-X$ direction \cite{EBOM:Sheng}, as shown in Fig. \ref{ebands}.

Compared with the eight-band ${\bf k}\cdot{\bf p}$ method, one of advantages of the EBOM is the absence of
spurious crossing between valence bands (see Fig. \ref{ebands}). To alleviate this deficiency in the ${\bf
k}\cdot{\bf p}$ method, terms proportional to $-(k_x^4 + k_y^4 + k_z^4)$ are proposed \cite{KP:Holma} to be
added to heavy-hole and light-hole bands to eliminate the resulting spurious solutions (see dash line in
Fig. \ref{ebands}). Although this modification does not alter effective masses at the $\Gamma$ point, it is
nevertheless seen to give unphysical dispersion of valence bands along the $\Gamma-X$ and $\Gamma-L$
directions, compared with that given by the EBOM.

When EBOM is applied to two different materials, a band offset between the unstrained island and matrix
material is needed to obtain all the fitting parameters.  The effect of strain is incorporated into EBOM
through the deformation potential Bir-Pikus's theory \cite{DP:Bir, EBOM:Sun} and through the piezoelectric
effect. An advantage of the EBOM over ETBM is its lack of ambiguity in fitting the deformation potential
parameters \cite{TB:Santoprete}.

\section{Strain distribution}

The strain in the vicinity of quantum dot is determined by the position of the quantum dot in the 
pyramid. Hence the domain of strain calculation should include the entire pyramid, with characteristic 
sizes on the micron scale. While the full atomistic and finite element calculation for the quantum dot 
and the pyramid will be implemented in the future, we report here calculations based on continuum 
elasticity theory \cite{STRAIN:Pryor}. We enclose the template in a rectangular computational box. A 
fixed boundary condition is applied to the base of the template while free-standing boundary conditions 
are implemented to all the other exposed facets. This is achieved by introducing dummy sites around the 
template where infinitesimal elastic constants are assigned, therefore, the force applied to the template 
facets can be neglected. The quality of results of continuum elasticity theory calculations are 
successfully verified by comparison with the results of atomistic valence-force-field calculations 
\cite{VFF:Groenen} for smaller structures.

The strain tensor $\epsilon$ is obtained by minimizing the following elastic energy functional
\begin{eqnarray}
E &=& \frac{1}{2} \int \biggl[ C_{11} \bigl(\epsilon_{xx}^2 + \epsilon_{yy}^2 + \epsilon_{zz}^2\bigl) + C_{44} 
\bigl(\epsilon_{xy}^2 + \epsilon_{yz}^2 + \epsilon_{zx}^2\bigl) \cr
&+& 2 C_{12} \bigl(\epsilon_{xx}\epsilon_{yy} + \epsilon_{yy}\epsilon_{zz} + \epsilon_{zz}\epsilon_{xx}\bigl) \biggl] 
d^3{\bf r},
\label{strainenergy}
\end{eqnarray}
where $C_{11}$, $C_{44}$, and $C_{12}$ are elastic constants of quantum dot material and  of InP pyramid material, 
the values of which are taken from Ref. \cite{KP:Holma}.

\begin{figure}
\vspace{5mm}
  \includegraphics[width=3.3in]{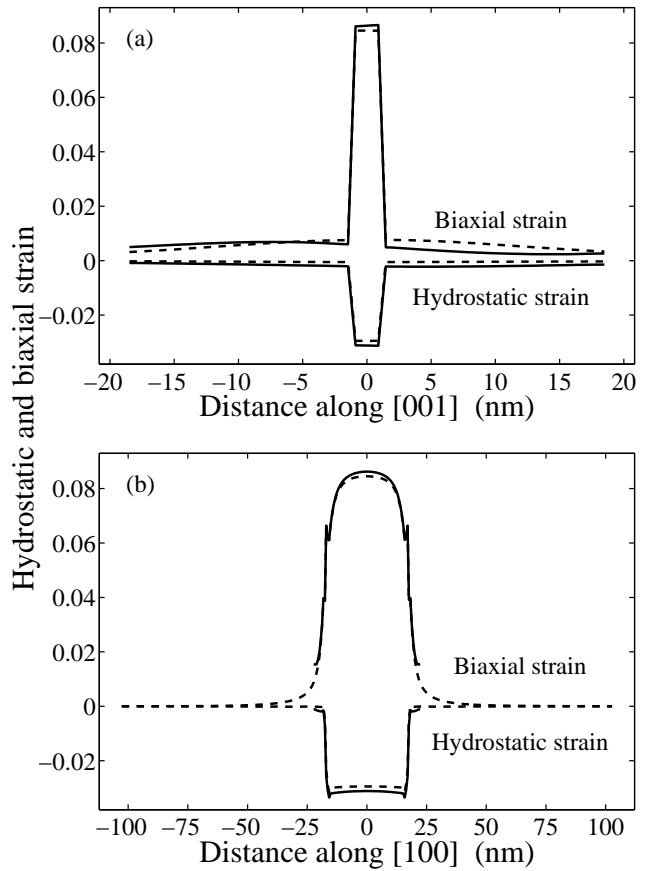}\\
  \caption{ Plot of hydrostatic and biaxial strain in the InAs/InP quantum dot along the [001] direction towards the
apex (a) and along [100] direction (b). For comparison,  strain profiles are shown in dash lines for a InAs dot of
the same geometry in an unpatterned InP substrate. }
\label{strain}
\end{figure}

Figure \ref{strain} shows two typical strain components, hydrostatic and biaxial strain, through the central
axis of the dot ([001] direction) and through the center of the dot along the base ([100] direction). The
hydrostatic and biaxial strain are defined \cite{KP:InAs/GaAs} by
\begin{eqnarray}
&H = \epsilon_{xx} + \epsilon_{yy} + \epsilon_{zz},& \cr \cr
&B^2 = (\epsilon_{xx} - \epsilon_{yy})^2 + (\epsilon_{yy} - \epsilon_{zz})^2 + (\epsilon_{zz} - \epsilon_{xx})^2& . 
\label{straintensor}
\end{eqnarray}
The former mainly affects the conduction bands while the latter only affects the valence bands, inducing the
splitting between the heavy-hole and light-hole band.

The same strain profiles are also shown for InAs dot in an unpatterned InP substrate. In this comparison we
assumed the same shape and size of the quantum dot, excluding the possibility that the unpatterned dots are
disk-like or lens-shaped. Hereafter, dot on unpatterned substrate is referred to as dot 1 and the dot on
patterned substrate is referred to as dot 2.

The comparison between them shows the effect of patterned substrate on the strain distribution. Along the
growth direction [see Fig. \ref{strain}(a)], both the hydrostatic and biaxial strain are found marginally
larger in dot 2 than in dot 1, which would lead to a small blue shift of emission energy. Along the
direction towards the apex, due to the physical limitation of the patterned substrate, obvious discrepancy
can be found between the strain profiles in the two dots that are in different environment. Much less
difference is found along the opposite direction where the patterned substrate becomes similar to bulk InP.
Along the [100] direction, the strain profiles are found similar for the two dots except for their slightly
different peak values. 

According to the deformation potential theory \cite{DP:Bir}, the local band edges are proportional to the
hydrostatic and biaxial strain. If we choose the valence band offset between unstrained InAs and InP as 400
meV \cite{EXP:Pettersson2000}, the depth of confinement along the growth direction is 420 meV and 530 meV
for electrons and holes, respectively, which is in agreement with previous calculation \cite{KP:Holma}. The
heavy-hole band is the top-most valence band, separated from the light-hole band by 150~meV due to the
biaxial strain. Although the band offset between unstrained InAs and InP is not well known, it has been
shown that the actual electronic structure are not sensitive to the value chosen in the calculation
\cite{TB:Santoprete}.

\section{Electronic structure}

The computational box for electronic calculation is chosen to be much smaller than that for the strain
calculation as we are interested only in the states confined in the dot. It contains the whole dot and some
sites outside the patterned substrate where much larger site energies are assigned. Its dimensions are
47.5 nm $\times$ 47.5 nm $\times$ 14.1 nm. In EBOM, each unit cell consists of four effective atoms, and
each atom has eight spin-orbitals.  In total, it gives rise to a large sparse matrix of dimension exceeding
five million. The electron states are obtained by solving the sparse matrix using the Lanczos algorithm
\cite{KP:InAs/GaAs}.

\begin{figure}
\vspace{5mm}
  \includegraphics[height=2.5in]{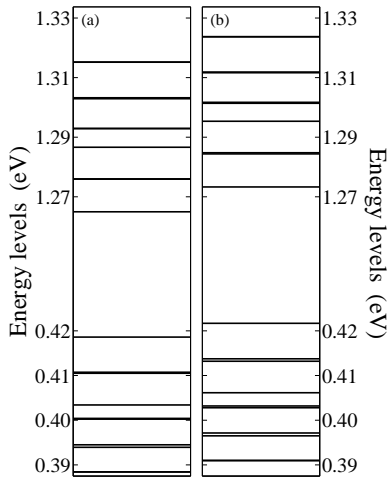}\\
  \caption{ Energy levels of the single InAs dot on unpatterned InP substrate (a) and on patterned substrate (b)
calculated by the EBOM. It is noted that the energy scale in conduction band is different from that in valence bands. }
\label{estates}
\end{figure}

Figure \ref{estates} (a) and (b) shows the calculated energy levels of electrons and holes by EBOM for dot 1
and 2, respectively. The calculated energy shells are characteristic to the square geometry of the dot, with
a single $s$-level, almost degenerate $p$-shell, and a d-shell consiting of three states: a
single  level followed by two degenerate levels. The order and degeneracies (1,2) of levels in the d-shell
differentiates the square dot from a lens-shaped (3) or disk-like quantum dot (2,1)
studied previously \cite{QD:Hawrylak}, and could be used to infer the shape from the optical spectra.

As can be seen, strain profile specific to nanotemplate (see Fig. \ref{strain}) induces shifts of energy
levels for the dot on patterned substrate. The shift in the conduction band, which is about 11 meV, is more
homogeneous than that in the valence bands (about 3 meV) as the latter is also under the influence of other
strain components.

\begin{figure}  
\vspace{5mm}
  \includegraphics[width=3.3in]{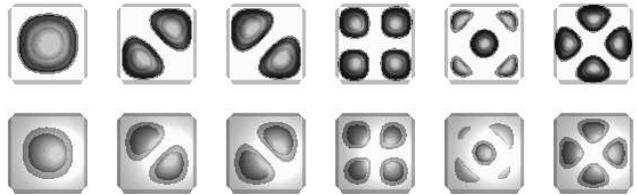}\\
  \caption{ Probability density of confined electron (upper row) and hole (lower row) states in the dot. }
\label{wavefun}
\end{figure}

The probability density of confined states is shown in Fig. \ref{wavefun}. The hole states are found
generally more confined than their electron counterparts. Also note that the electron and hole states of
p-shell are oriented in the opposite directions. This is a result of piezoelectric potential which attracts
electrons while simultaneously repelling holes.

We can label the $p$-like states according to the extension direction of their wave functions. In the
conduction band, the two $p$-like states are $p_+^e$ and $p_-^e$, localized along $[110]$ and $[1\bar{1}0]$
directions, with energies $E_{p_+^e}<E_{p_-^e}$. In the valence bands, the two $p$-like states, $p_-^h$ and
$p_+^h$, have a different order, i.e. $E_{p_+^h}<E_{p_-^h}$ \cite{note}. 

This inverted order of energy of states with similar symmetry is due to the enhanced piezoelectric effect 
in the dot on patterned substrate. The piezoelectric potential is calculated from the piezoelectric 
charge density as given by
\cite{KP:InAs/GaAs}
\begin{equation}
\rho_P({\bf r}) = -2e_{14} \Bigl[ \frac{\partial\epsilon_{yz}({\bf r})}{\partial x} + \frac{\partial\epsilon_{xz}({\bf
r})}{\partial y} + \frac{\partial\epsilon_{xy}({\bf r})}{\partial z} \Bigl] ,
\label{piezocharge}
\end{equation}
where $e_{14}$ is the piezoelectric constant. Unlike InAs/GaAs dots where $e_{14}$ in GaAs is four times 
larger than that in InAs, $e_{14}$ in InP is smaller than in InAs.

\begin{figure}  
\vspace{5mm}
  \includegraphics[width=3.3in]{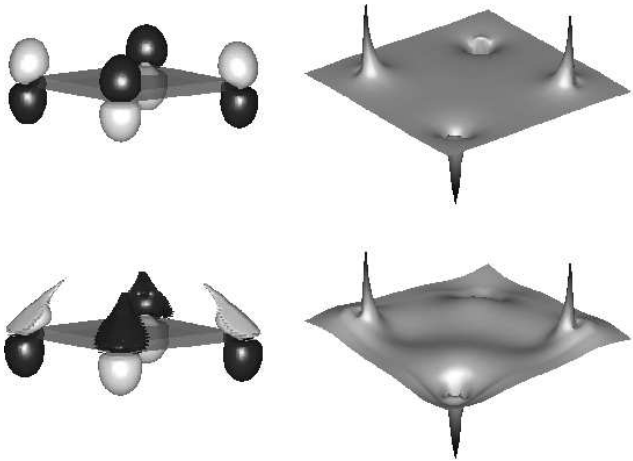}\\
  \caption{ Effect of nanotemplate : isosurface plots of piezoelectric charge density in the dot on patterned substrate
(lower-left plot) and the one on unpatterned substrate (upper-left plot). Bright (dark) grey area has positive
(negative) charge. The right plots show the corresponding piezoelectric potentials as seen by an electron along the
quantum-dot layer which is about 0.6 nm above the bottom of dots. The peak values of both potentials are about $\pm8$
meV. }
\label{piezopot}
\end{figure}

The left two plots in Figure \ref{piezopot} show the distribution of piezoelectric charge density for the 
two dots in different environment. The piezoelectric charge in dot 1 is seen to have a nearly symmetric 
distribution, i.e., a part of positive charge appears in pair with another one of negative charge. The 
two parts are almost symmetric and hence compensate their effect on electron and hole states, which can 
be seen in the corresponding piezoelectric potential shown in the upper-right panel. The potential is 
found to be well localized around the four corners of the dot. Because the positive charge is paired with 
the negative charge, the potential is very similar to that of dipoles, and is almost zero inside the 
dot. This leads to only 0.1 meV splitting between the $p_+^e$ and $p_-^e$ states.

In dot 2, the piezoelectric charge density is seen to be localized below the dot as well as along the facets
of the substrate. This asymmetric distribution of the charge density induces a very different piezoelectric
potential. As seen in the lower-right part of Fig. \ref{piezopot}, the potential inside the dot has
significant value. This leads to a splitting of about 0.5 meV between $p_+^e$ and $p_-^e$, and inverts the
order of states in valence bands.

\begin{figure}  
\vspace{5mm}
  \includegraphics[width=3.3in]{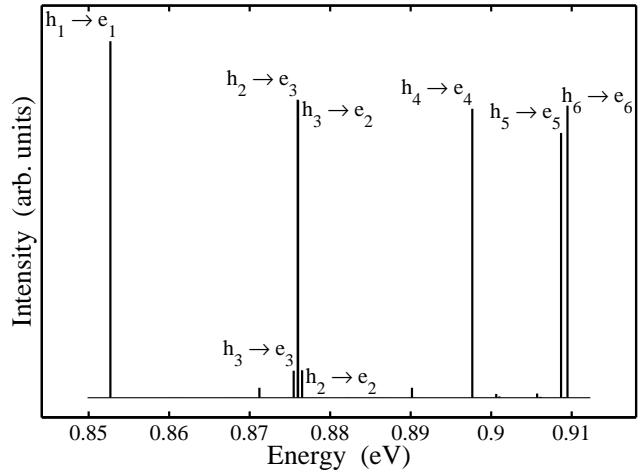}\\
  \caption{ Intensity of interband transitions between hole and electron states. }
\label{interpl}
\end{figure}

Figure \ref{interpl} shows the intensity of interband transitions (joined optical density of states)
\cite{KP:InAs/GaAs} between the calculated hole and electron states. Here, we only plot the transitions
polarized within the plane perpendicular to the growth direction, which corresponds to those from the
heavy-hole part of the hole states. After the ground state transition, the next two strong transitions are
those between the $p$-like hole and electron states. The transitions between electron and hole states that
have different symmetry, i.e., $p_-^h \rightarrow p_+^e$ and $p_+^h \rightarrow p_-^e$, are seen to have
lower intensity.

A further calculation shows that the tight-binding results can be fairly well approximated by a 
single-band effective-mass approach \cite{EBOM:Sheng} using the following effective mass parameters, 
$m_e^* = 0.065$, $m_h^{001} = 0.32$, and $m_h^{110} = 0.17$. The comparison of these parameters with 
those of the bulk materials exhibits the effect of strain on the effective mass renormalization in 
quantum dots. For example, the electron effective mass is enhanced from 0.024 in bulk InAs to 0.065, 
which is close to the effective mass in bulk InP (0.077). The anisotropy in the hole effective mass 
tensor is reversed, the holes become much lighter (comparing with 0.639 in bulk InAs and 0.933 in bulk 
InP) within the plane perpendicular to the growth direction. This results in comparable energy 
separations in conduction (14.8 meV) and valence (8.0 meV) bands.

\section{Emission spectra of dots on patterned and unpatterned substrates }

\begin{figure}  
\vspace{5mm}
  \includegraphics[width=3.3in]{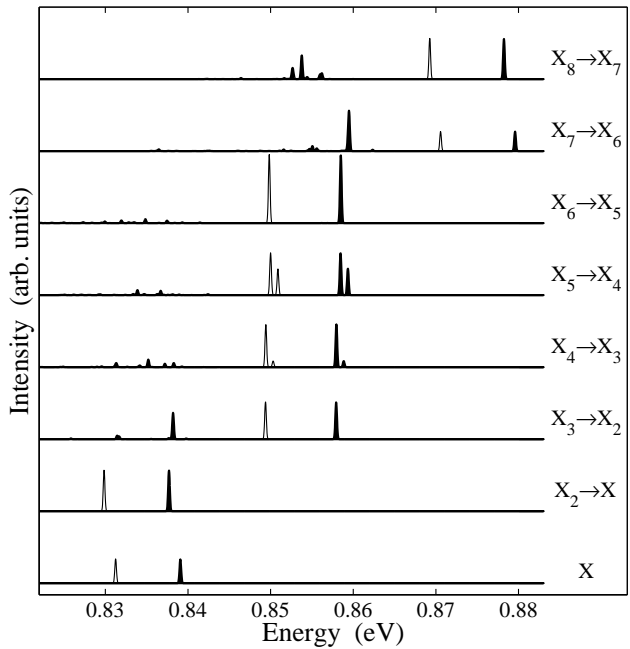}\\
  \caption{ Emission spectrum from individual multi-exciton complex obtained by solving the many-body
Hamiltonian using the configuration-interaction method, shown as peaks filled in dark color. For comparison,
the results for the dot on an unpatterned substrate are shown as unfilled peaks. }
\label{mexpl}
\end{figure}

We now turn to investigate the effect of nano-template growth on electronic states as could be observed from
emission spectra as a function of the population of the quantum dot with electrons and holes. The
interaction of electrons and holes significantly changes the emission spectra from those expected from the
joint optical density of states. The electrons and hole interaction is described via the following
Hamiltonian,

\begin{eqnarray}
H_{ex} &=& \sum_i E_i^e c_i^+ c_i + \sum_i E_i^h h_i^+ h_i -
           \sum_{ijkl} V^{he}_{ijkl} h_i^{+} c_j^{+} c_k h_l \cr
       &+& \frac{1}{2} \sum_{ijkl} V^{ee}_{ijkl} c_i^+ c_j^+ c_k c_l + 
           \frac{1}{2} \sum_{ijkl} V^{hh}_{ijkl} h_i^+ h_j^+ h_k h_l ,
\label{excitons}
\end{eqnarray}
where $E_i^e$ and $E_i^h$ are the energy levels shown in Fig. \ref{estates}, and $V^{ee}$, $V^{hh}$, and
$V^{he}$ are the Coulomb matrix elements which can be calculated using the single-particle states as shown
in Fig. \ref{wavefun}. In the calculation reported below, the multi-exciton configurations are built from the
first twelve electron and twelve hole single-particle states (with quasi-spin). These configurations
describe the lowest $s$, $p$, and $d$ shells. The Hamiltonian is then expanded on these configurations and
solved by the configuration-interaction method. For more detailed description of this calculation we refer
to our previous work \cite{EBOM:Sheng}.

Figure \ref{mexpl} shows the emission spectra for dot 1 and 2 as a function of photon energy and 
increasing number of excitons . The emission intensity is calculated for all the possible transitions 
between the n-exciton complex and (n-1)-exciton complex at a temperature of 4.2~K. The comparison of the 
spectra between the two dots should show the effect of nano-template, in particular the effect of 
inverted order of states in the conduction and valence band of the dot on patterned substrate. The 
inverted order of $p$-like states in valence bands is due to different strain distribution, especially in 
the region close to the corners of the dot structure, and hence different piezoelectric potential. 
However, we find very little difference in the spectra between the two dots except for an overall blue 
shift which reflects the difference found in the corresponding single-particle energy spectra (see Fig. 
\ref{estates}).

\begin{figure}  
\vspace{5mm}
  \includegraphics[width=3.3in]{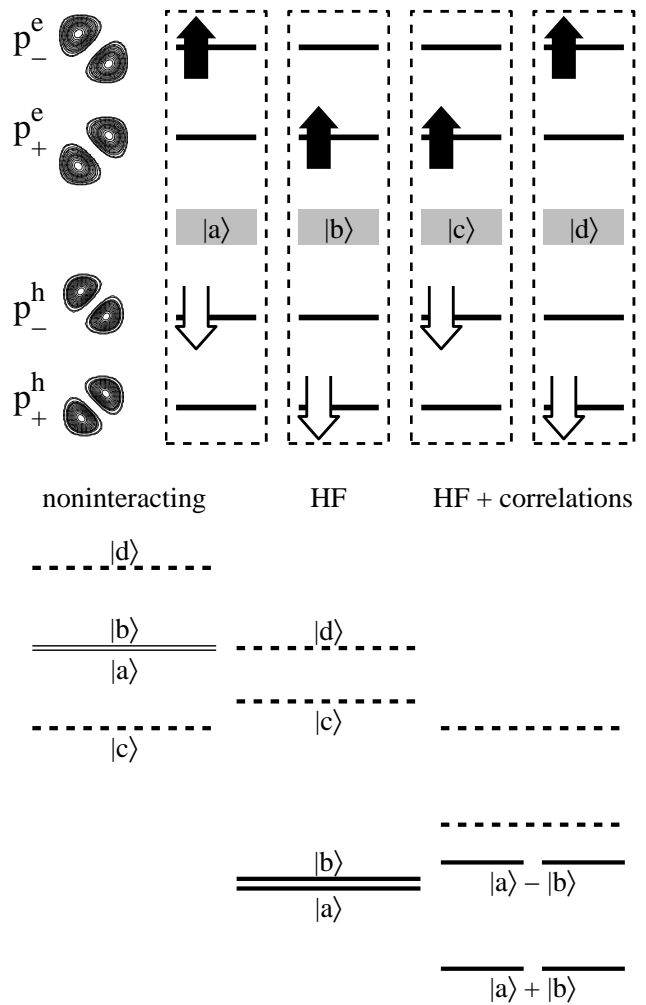}\\
  \caption{ Upper panel: Four possible configurations of the $p$-shell states whose probability densities are 
plotted in the left, Lower panel: Schematic energy levels. }
\label{triexciton}
\end{figure}

Let us first look at tri-exciton which is the first non-trivial case. Fig. \ref{triexciton} illustrates 
four configurations of the $p$-shell states for a tri-exciton with definite spin projection in its upper 
panel. The ground state of the non-interacting tri-exciton complex is given by the configuration $|c 
\rangle = |p_-^h\rangle |p_+^e\rangle$ as it has the lowest kinetic energy.  The other two configurations 
$|a \rangle = |p_-^h\rangle |p_-^e\rangle$ and $|b \rangle = |p_+^h\rangle |p_+^e\rangle$ have higher 
kinetic energies due to the splittings between the $p_+^{e(h)}$ and $p_-^{e(h)}$ states. Hence the ground 
state of the non-interacting tri-exciton would be dominated by $|c \rangle$, and is not be optically 
active. The energies of tri-exciton configurations and a schematic representation of their oscillator 
strength (solid line - high, dashed line - low )  are shown in the lower panel of Fig. \ref{triexciton}.

The non-interacting tri-exciton ground state is the dark configuration $|c \rangle $.  However, the order 
of energy levels changes once electron-hole attraction (Hartree-Fock energy) in each configuration is 
included.  As the spatial overlap between $p_-^h$ and $p_+^e$ is much smaller (see Fig. \ref{estates})  
than either that between $p_-^h$ and $p_-^e$, or that between $p_+^h$ and $p_+^e$, the corresponding 
Coulomb interaction is smaller by $2.3$ meV compared with those of the other two pairs. Taking this into 
account, configuration $|c \rangle$ has higher Hartree-Fock energy than $|a \rangle$ and $|b \rangle$ by 
about $1.8$ meV. Configuration $|d \rangle$ has an even higher energy than $|c \rangle$ because of the 
splitting between the $p_+^{e(h)}$ and $p_-^{e(h)}$ states. This arrangement of HF energy levels is shown 
in lower-central panel of Fig. \ref{triexciton}. Once correlation are included ( lower-right panel of 
Fig. \ref{triexciton} ), the ground state of tri-exciton primarily consists of a symmetric combination of 
configurations $|a \rangle$ and $|b \rangle$ \cite{CI:Hawrylak}, and becomes optically active.

When the states beyond the second-shell are included, the ground state of four-exciton complex ($X_4$) is
dominated by the singlet-singlet (SS) configuration \cite{CI:Hawrylak} where the two electrons (holes)
occupying the two $p$-like states have anti-parallel spins.  However, a state dominated by a triplet-triplet
(TT) configuration is lying very close to the ground state. Hence, there are two emission peaks found in the
p-shell emission spectrum of 4X. Depending on the temperature and spin relaxation, the emission from this
excited state may become stronger. The ground state of five-exciton complex is well isolated from excited
states. The two emission peaks found in the $p$-shell are the result of recombinations to the ground and
first excited 4X states. The emission spectrum from $s$ and $p$ shells for dot on patterned substrate is
very similar to the dot on unpatterned substrate. Hence patterning did not deteriorate optical properties.  

\subsection{Power-dependent emission}

In the experiment \cite{EXP:Williams}, the emission spectra are measured as a function of excitation power
and include the contribution from all the multi-exciton complexes. In our calculation, we solve the rate
equations \cite{EXP:Dekel} for given excitation power to obtain the distribution probability of each
multi-exciton complex. The overall emission spectra are computed by the summation of the contribution from
individual multi-exciton complex weighted by the calculated probabilities.

\begin{figure}  
\vspace{5mm}
  \includegraphics[width=3.3in]{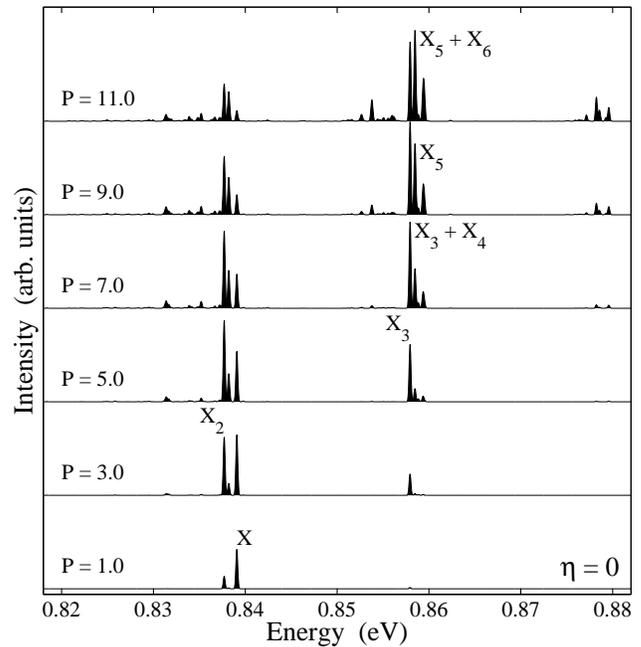}\\
  \caption{ The calculated photoluminescence spectra at various excitation power P (in arbitrary units). The major 
individual emission peaks are labeled according to their origin. }
\label{plspec}
\end{figure}

Figure \ref{plspec} shows the dependence of photoluminescence spectrum on the excitation power controlled 
by dimensionless parameter $P$. The first emission line appearing in the $s$ shell is from the single 
exciton and the second line is from a bi-exciton. The calculated single exciton emission line is about 15 
meV below what the experiment measured. The calculated bi-exciton binding energy (measured by the 
separation between the $X$ and $X_2$ lines) is 1.4~meV, which is larger than what the experiment measured 
(0.9~meV).

In the $s$ shell between the $X$ and $X_2$ lines lies an emission line from the tri-exciton. As the 
excitation power increases, the emission from the $p$ shell is seen to have larger intensity than the $s$ 
shell. In the $p$ shell, the major emission peaks from $X_3$ ($X_5$) and $X_4$ ($X_6$) overlap each other 
due to their very similar emission energies(see also Fig. \ref{mexpl}). In addition, emission from 
higher-lying states beyond the $p$-shell can also be seen at higher energies, which is characteristic to 
that from a single $2s$ level of the d-shell, populated by one and two electron-hole pairs, much like an 
s-shell.

The energy separation between the emission lines of the single exciton and tri-exciton defines the
separation of the $s$ and $p$ shell. Compared with the experimental value \cite{EXP:Williams}, which is
about 18.0 meV, our calculation gives 17.9 meV. This shell separation is largely determined by the lateral
size of the dot and not sensitive to the height of the dot.

\section{Tuning optical properties with nanotemplate}

We now turn to the possibility of tuning the shape of a quantum dot with the shape of the template. We
assume that the change in the template leads to the controlled elongation of the quantum dot. The degree of
elongation is defined as $\eta = (d-d^\prime)/d$ where $d$ and $d^\prime$ denote the dimensions of the
rectangular base of the elongated dot. We keep $d$ as a constant (36 nm) and change $d^\prime$ to obtain
different elongated geometries.

\begin{figure}  
\vspace{5mm}
  \includegraphics[width=3.3in]{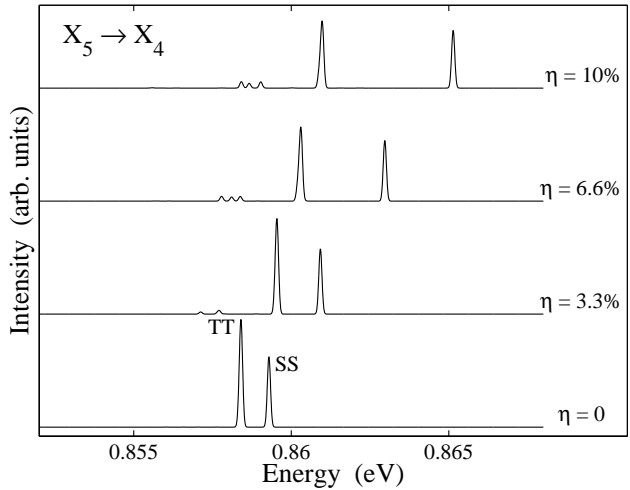}\\
  \caption{ Emission spectra of the five-exciton complex in quantum dots with increasing elongation}
\label{mexple}
\end{figure}

To explain the effect of elongation on the emission spectrum we discuss here in detail the emission from the
five-exciton complex. Even without elongation the emission from the 5X complex results in two emission lines
in the $p$-shell, as shown in Fig.\ref{mexpl}. These two lines originate from two different 4X states:
singlet-singlet (SS) and triplet-triplet (TT) 4X states \cite{CI:Hawrylak}. The splitting between the SS and
TT states is mainly determined by the competition between exchange energy of triplets and correlations in
singlet-singlet configurations. When there is elongation in the geometry of the dot, the almost degenerate
single-particle states in the $p$-shell have an extra splitting proportional to the elongation. This
splitting is present in the splitting of the SS and TT states of 4X complex, which can be seen in the
emission spectrum of 5X.
This is illustrated in Fig. \ref{mexple} which shows the dependence of emission spectra on the elongation of
the structure.

The splitting between the two emission peaks of 5X is seen to increase with the degree of elongation, which
can be well explained by the increasing splitting between the single-particle states in the $p$-shell. The
emission peaks of all the multi-exciton complexes are found to have blue shifts as the degree of elongation
increases, which is due to the shrinking of the volume of the quantum dot.

\begin{figure}  
\vspace{5mm}
  \includegraphics[width=3.3in]{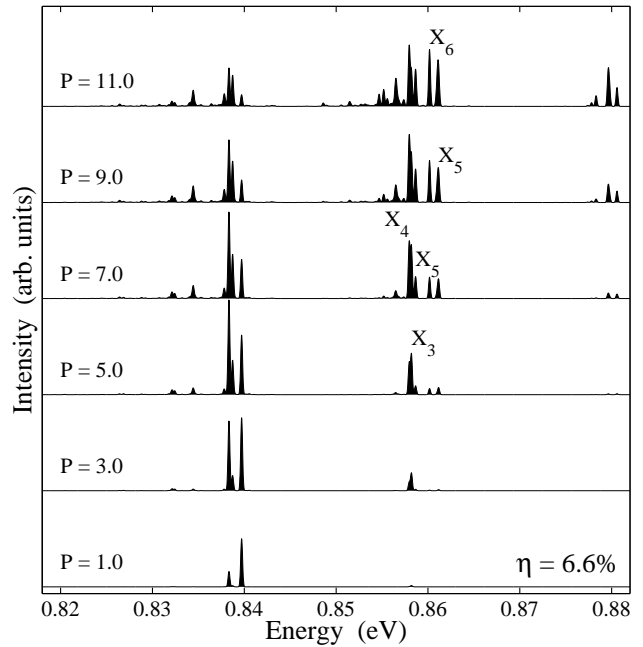}\\
  \caption{ The photoluminescence spectra at various excitation power P for an elongated dot 
($\eta = 6.6\%$).}
\label{plspece}
\end{figure}

We summarize the effect of the elongation on the emission spectra as a function of excitation power in Fig.  
\ref{plspece} which shows the emission spectra for an elongated dot with $\eta = 6.6\%$.  While the
elongation is seen not to affect the emission in the $s$-shell, the splittings it induced in the $p$-shell
can be clearly identified. The order of individual emission peaks in the $p$-shell is also found sensitive
to the elongation.

\section{Conclusions}

In conclusion, we have presented a theoretical study of electronic and optical properties of a single InAs
dot on InP patterned substrate. We have calculated the strain distribution in nanotemplate with appropriate
boundary conditions. By applying the atomistic tight-binding effective-bond-orbital method, we have obtained
the electronic structure of the dot on patterned substrate. By comparing the results with those for the dot
on unpatterned substrate, we have shown that the patterned substrate breaks the reflection symmetry of the
structure, enhances the piezoelectric effect, and leads to the inverted order of the $p$-shell states. We
have shown that Coulomb interactions restore the order of electron hole states and result in emission
spectra unaffected by the patterning process. We have identified features in the emission spectra which can
be related to the shape of the quantum dots, in particular the characteristic emission pattern of the d-shell.
 We have explored the possibility of tuning the 
emission spectra by changing the shape of the quantum dot in a controlled way using
the nano-template.  The theoretical results support the notion
that the nanotemplate growth of single self-assembled quantum dots is a promising route for the control of
the position of self-assembled dots, and hence enables their increased functionality when combined with
cavities, magnetic ions, doping and gates. In turn, it is hoped that better experimental control will allow
a systematic comparison of microscopic theories with experiment.

\acknowledgments
This work is supported by the NRC HPC project, NRC-HELMHOLTZ grant, and Canadian Institute for Advanced Research. 
The authors would like to thank R. L.
Williams and G. C. Aers for discussions.


\begin{thebibliography}{}

\bibitem{QD:Jacak}
Lucjan Jacak, Pawel Hawrylak, and Arkadiusz Wojs,
{\it Quantum dots}, (Springer, Berlin, New York, 1998).

\bibitem{QD:Bimberg}
D. Bimberg, M. Grundmann, and N. N. Ledentsov,
{\it Quantum Dot Heterostructures}, (John Wiley \& Sons, UK, 1998).

\bibitem{QD:Hawrylak}
P. Hawrylak and M. Korkusinski,
{\it Electronic properties of self-assembled quantum dots}, in {\it Single Quantum dots},
edited by P. Michler (Springer, Berlin, New York, 2003).

\bibitem{QD:PS}
% Seeded self-assembled GaAs quantum dots grown in two-dimensional V grooves by selective metalorganic chemical-vapor ...
S. Ishida, Y. Arakawa, and  K. Wada,
Appl. Phys. Lett. {\bf 72}, 800 (1998).

\bibitem{EXP:Williams}
% Quantum dot site-selection using in situ prepared nano-templates
R. L. Williams, G. C. Aers, J. Lefebvre, P. J. Poole and D. Chithrani,
Physica E {\bf 13}, 1200 (2002).

\bibitem{EXP:Chithrani}
% Optical spectroscopy of single, site-selected, InAs InP self-assembled quantum dots
D. Chithrani, R. L. Williams, J. Lefebvre, P. J. Poole, and G. C. Aers,
Appl. Phys. Lett. {\bf 84}, 978 (2004).

\bibitem{EXP:Dalacu}
% InAs/InP quantum-dot pillar microcavities using SiO[2]/Ta[2]O[5] Bragg reflectors with emission around 1.55  m
Dan Dalacu, Daniel Poitras, Jacques Lefebvre, Philip J. Poole, Geof C. Aers, and Robin L. Williams, 
Appl. Phys. Lett. {\bf 84}, 3235 (2004).

\bibitem{EXP:Kim}
% proposal of single photons from quantum dots
J. Kim, O. Benson, H. Kan, and Y. Yamamoto, Nature (London) {\bf 397}, 500 (1999).

\bibitem{EXP:Benson}
% proposal of entangled photon pairs
O. Benson, C. Santori, M. Pelton, and Y. Yamamoto, Phys. Rev. Lett. {\bf 84}, 2513 (2000).

\bibitem{KP:Holma}
% Calculations of the electronic structure of strained InAs quantum dots in InP
M. Holma, M.-E. Pistol, and C. Pryor,
J. Appl. Phys. {\bf 92}, 932 (2002).
% correction to k*p Hamiltonian for InAs/InP

\bibitem{EBOM:Chang}
% Bond-orbital models for superlattices
Yia-Chung Chang,
Phys. Rev. B {\bf 37}, 8215 (1988).
% original proposition of EBOM

\bibitem{EBOM:Sun}
% Modeling self-assembled quantum dots by the effective bond-orbital method
Sophia J. Sun and Yia-Chung Chang,
Phys. Rev. B {\bf 62}, 13631 (2000).
% application in SADs calculation

\bibitem{DP:Bir}
% deformation potential theory
G. L. Bir and G. E. Pikus, {\it Symmetry and Strain-Induced Effects in Semiconductors}, (Wiley, New York, 1974).

\bibitem{EXP:McCaffrey}
% Interpretation and modeling of buried InAs quantum dots on GaAs and InP substrates
J. P. McCaffrey, M. D. Robertson, P. J. Poole, B. J. Riel, and S. Fafard, 
J. Appl. Phys. {\bf 90}, 1784 (2001).

\bibitem{EXP:Pettersson2000}
% Electrical and optical properties of self-assembled InAs quantum dots in InP studied 
% by space-charge spectroscopy and photoluminescence
H. Pettersson, C. Pryor, L. Landin, M.-E. Pistol, N. Carlsson, W. Seifert, and L. Samuelson,
Phys. Rev. B {\bf 61}, 4795 {2000}.
% luminescence from the ground state : 737 meV

\bibitem{TB:Bryant}
% Electron-hole correlations in semiconductor quantum dots with tight-binding wave functions
S. Lee, L. Jonsson, J. W. Wilkins, G. W. Bryant, and G. Klimeck,
Phys. Rev. B {\bf 63}, 195318 (2001).
% Tight-binding calculation of single-particle states inself-assembled quantum dots

\bibitem{EBOM:Loehr}
% Improved effective-bond-orbital model for superlattices
John P. Loehr,
Phys. Rev. B {\bf 50}, 5429 (1994)
% second-nearest neighbor model

\bibitem{EBOM:Meyer}
% Curie-temperature enhancement in ferromagnetic semiconductor superlattices
I. Vurgaftman and J. R. Meyer,
Phys. Rev. B {\bf 64}, 245207 (2001).
% application of EBOMs to ferromagnetic semiconductor superlattices

\bibitem{KP:InAs/GaAs}
H. Jiang and J. Singh, Phys. Rev. B {\bf 56}, 4696 (1997); 
Craig Pryor, Phys. Rev. B {\bf 57}, 7190 (1998);
O. Stier, M. Grundmann, and D. Bimberg, Phys. Rev. B {\bf 59}, 5688 (1999); 
Weidong Sheng and J.-P. Leburton, Appl. Phys. Lett. {\bf 80}, 2755 (2002).
% application of k*p to InAs/GaAs dots

\bibitem{TB:Klimeck}
% sp3s* Tight-binding parameters for transport simulations in compound semiconductors
G. Klimeck, R. C. Bowen, T. B. Boykin, T. A. Cwik, 
Superlattices and Microstructures, {\bf 27}, 519 (2000).
% experimental band structure

\bibitem{EBOM:Sheng}
% Multiband theory of multi-exciton complexes in self-assembled quantum dots
Weidong Sheng, S.-J. Cheng, Pawel Hawrylak,
Phys. Rev. B, {\bf 71}, 035316 (2005).
% multiband theory of excitons

\bibitem{TB:Santoprete}
% Tight-binding study of the influence of the strain on the electronic properties of InAs/GaAs quantum dots
R. Santoprete, Belita Koiller, R. B. Capaz, P. Kratzer, Q. K. K. Liu, and M. Scheffler,
Phys. Rev. B {\bf 68}, 235311 (2003).

\bibitem{STRAIN:Pryor}
% Comparison of two methods for describing the strain profiles in quantum dots
C. Pryor, J. Kim, L. W. Wang, A. J. Williamson, and A. Zunger,
J. Appl. Phys. {\bf 83}, 2548 (1998).

\bibitem{VFF:Groenen}
% Strain distribution and optical phonons in InAs/InP self-assembled quantum dots
J. Groenen, C. Priester, and R. Carles,
Phys. Rev. B {\bf 60}, 16013 (1999).
% valence-force-field method applied to 3 nm high and 25 nm wide dots or 7 nm high and 45 nm wide dots
% InAs/InP islands are in fact slightly elongated along [1-10]

\bibitem{note}
The energy of hole states is measured from the top of valence bands in bulk GaAs.

\bibitem{CI:Hawrylak}
% Excitonic artificial atoms: Engineering optical properties of quantum dots
Pawel Hawrylak,
Phys. Rev. B {\bf 60}, 5597 (1999).
% semi-analytical calculation of multi-exciton for parabolic dots

\bibitem{EXP:Dekel}
% Carrier-carrier correlations in an optically excited single semiconductor quantum dot
E. Dekel, D. Gershoni, E. Ehrenfreund, J. M. Garcia, and P. M. Petroff,
Phys. Rev. B {\bf 61}, 11009 (2000).
% rate equation 

\end{thebibliography}
\end{document}